\documentclass[aps, pra, a4paper,showpacs,twocolumn,10pt]{revtex4-1}
\usepackage{bbm, amsmath, amssymb, amsthm, bm,textcomp, nicefrac,geometry,ragged2e}%, subcaption}
\usepackage{graphicx,epstopdf,color,verbatim,enumitem}
\usepackage[dvipsnames]{xcolor}

\usepackage[bbgreekl]{mathbbol}
\usepackage{graphicx,epstopdf,color,verbatim,enumitem,ulem}

\definecolor{myurlcolor}{rgb}{0,0,0.7}
\definecolor{myrefcolor}{rgb}{0.8,0,0}
\usepackage[unicode=true,pdfusetitle, bookmarks=false,bookmarksnumbered=false,
bookmarksopen=false, breaklinks=false,pdfborder={0 0 0},backref=false,
colorlinks=true, linkcolor=myrefcolor,citecolor=myurlcolor,urlcolor=myurlcolor]
{hyperref}

\usepackage{pbox,hyperref,array}
\usepackage[caption=false]{subfig}

\newcommand{\eins}{\mathbbm{1}}

\newcommand{\ket}[1]{\left|#1\right\rangle}

\newcommand{\bra}[1]{\left\langle #1\right|}

\newcommand\ketbra[1]{|#1\rangle\langle#1|}

\newcommand{\be}{\begin{equation}}
\newcommand{\ee}{\end{equation}}
\definecolor{darkorange}{RGB}{255,140,0}

\newcommand{\bea}{\begin{eqnarray}}
\newcommand{\eea}{\end{eqnarray}}

\makeatletter
\newtheorem*{rep@theorem}{\rep@title}
\newcommand{\newreptheorem}[2]{%
\newenvironment{rep#1}[1]{%
 \def\rep@title{#2 \ref{##1}}%
 \begin{rep@theorem}}%
 {\end{rep@theorem}}}
\makeatother

\newreptheorem{theorem}{Theorem}

\newtheorem*{result*}{Result}

\graphicspath{{./figures/}}
\DeclareGraphicsExtensions{.png,.pdf,.eps}

\begin{document}
\title{Two-dimensional quantum repeaters}
\author{J. Walln\"ofer$^1$, M. Zwerger$^2$, C. Muschik$^{1,3}$, N. Sangouard$^2$ and W.~D\"ur$^1$}
\affiliation{$^1$ Institut f\"ur Theoretische Physik, Universit\"at Innsbruck, Technikerstr. 21a, A-6020 Innsbruck,  Austria\\
$^2$ Department of Physics, University of Basel, Klingelbergstrasse 82, 4056 Basel, Switzerland\\
$^3$ Institute for Quantum Optics and Quantum Information of the Austrian Academy of Sciences, 6020 Innsbruck, Austria}

\date{\today}

\begin{abstract}
The endeavour to develop quantum networks gave rise to a rapidly developing field with far reaching applications such as secure communication and the realisation of distributed computing tasks. This ultimately calls for the creation of flexible multi-user structures that allow for quantum communication between arbitrary pairs of parties in the network and facilitate also multi-user applications. To address this challenge, we propose a 2D quantum repeater architecture to establish long-distance entanglement shared between multiple communication partners in the presence of channel noise and imperfect local control operations.
The scheme is based on the creation of self-similar multi-qubit entanglement structures at growing scale, where variants of entanglement swapping and multi-party entanglement purification are combined to create high fidelity entangled states. We show how such networks can be implemented using trapped ions in cavities.
\end{abstract}
\pacs{03.67.Hk, 03.67.Lx, 03.67.-a}
\maketitle

%%%%%%%%%%%%%%%%%%%%%%%%%INTRODUCTION%%%%%%%%%%%%%%%%%%%%%%%%%%%
\section{Introduction}

Long-distance quantum communication is one of the most appealing applications of quantum technology. It promises secure classical communication via quantum key distribution and is also essential for distributed quantum computation. High-rate quantum communication over long distances is possible using quantum repeaters, which either employ quantum error correction \cite{Knill96,Zw14,Muralidharan2014} or create long-distance entanglement between two parties from shorter-distance entanglement via swapping and processing operations \cite{Br98, Sa11} (see also \cite{Duan:2001aa,ladd2006,Loock2006,Childress2006,Jiang2007,Collins2007,Silvestre2013,Azuma2015,Pirandola2016,Azuma2016}), thereby overcoming limitations due to noisy and lossy channels and limited local control.

However, in a real-world application such as a quantum internet \cite{Kimble08}, one deals with a multi-user communication network. In such a network, the goal is not only to establish long-distance entangled pairs between fixed communication partners. One rather demands a flexible structure, where any given pair of parties can share entanglement and communicate, and that multipartite entangled states can be established between various communication partners to enable multi-user applications \cite{Ep15}. The latter is of particular importance for applications beyond two party quantum cryptography, for instance in distributed quantum computation \cite{Beals20120686}, secret voting and secret sharing \cite{Hillery05}, clock synchronization \cite{Komar2014}, or remote sensing. Depending on the task, certain kinds of entangled states need to be generated.

Here we address this inherently two-dimensional problem with a 2D-strategy. More specifically, we generalize the idea of the quantum repeater to 2D networks and propose an architecture that enables the direct generation of different kinds of entangled multi-party states \cite{RevModPhys.81.865} that are required for the applications mentioned above over large distances and between arbitrary communication partners, including Greenberger-Horne-Zeilinger (GHZ) states and 2D cluster states \cite{Briegel2001}.
As we show below, our direct 2D approach offers --in certain parameter regimes-- an advantage over a combination of 1D networks, where multipartite entangled states are established by appropriately combining bipartite entangled pairs. It can tolerate more errors, reaches higher fidelities and requires fewer local resources for storage.

More precisely, we show how to establish entangled states of fixed kind and size on larger and larger scales. The procedure combines several elementary (short-distance) states to obtain an equivalent state, but at larger distance. Imperfections in state preparation and local operations lead to a limited fidelity, which can be resolved by using multiparty entanglement purification (MEP) \cite{Du03,Du05,Kr06a,Du07,Glancy2006} to re-establish states with high fidelity from several copies. This repeater cycle can be applied in a nested way, similar as in the 1D repeater \cite{Br98} leading to states of arbitrary distance on the 2D lattice. By combining states of different scale, one obtains a multi-user communication network where all parties can participate and share entanglement. We illustrate this approach using 3-party GHZ states on a triangular lattice, and 8-party 2D cluster states on a rectangular lattice. Apart from the standard operational mode described above (mode I), where entangled states with a fixed number of parties are distributed over long distances, we also consider a variant of the scheme (mode II), where entangled states of growing size, i.e. with a larger number of parties involved, are generated among the 2D network. In this way one can e.g. produce a distributed 2D cluster state shared among all parties of the network. This state can then be used to establish pairwise quantum communication channels, but also for distributed measurement-based quantum computation \cite{Raussendorf2001,Briegel2009}.

The proposed scheme can make use of existing or currently developed platforms for 1D communication networks, as the experimental requirements are essentially the same. Only at the lowest level, the production of entangled pairs needs to be replaced by the preparation of GHZ states. For concreteness, we analyze the performance of an implementation using trapped ions in cavities for realistic noise parameters and show that appealing entangled states can be distributed over a thousand kilometers using present-day or near-future technologies.

The paper is organized as follows. In section \ref{sec:architecture} we introduce different schemes to distribute long-distance multipartite entanglement using a 2D approach. We mainly discuss
methods based on GHZ states and 2D cluster states, but the approach is not exclusive to these states. In the remaining sections we focus on a particular
protocol based on three qubit GHZ states that allows one to establish a two-dimensional communication network. In section \ref{sec:analysis} we analyse this 2D protocol and determine the errors it can tolerate.
In section \ref{sec:measurementbased} we consider a measurement based implementations of the scheme. In section \ref{sec:1dcompare} we give a
comparison between our 2D approach and networks based on 1D repeaters. In section \ref{ions} we discuss a concrete physical implementation based on trapped ions of our 2D repeater scheme without entanglement purification, and compare the performance to 1D strategies. We summarize our findings and conclude in section \ref{sec:summary}, while some technical details and additional
results regarding the use of different MEPs can be found in the appendix.

%%%%%%%%%%%%%%%%%%%%%%%%%%BACKGROUND%%%%%%%%%%%%%%%%%%%%%%%%%%%

\section{2D repeater architecture \label{sec:architecture}}
We consider a regular 2D network, where the communication parties are located at the nodes of the lattice and are connected by quantum channels.  %that are depicted by edges.
The goal is to establish high-fidelity long-distance entangled states shared between multiple communication partners. This is achieved by using entangled states generated over short distances, and connecting and purifying them by means of local operations.

%The standard operational mode, mode I, is to generate a certain type of entangle state, e.g. a $m$-qubit GHZ state
\subsection{Standard operational mode I}
\begin{figure}[ht]
 \centering
 \includegraphics[width=\columnwidth]{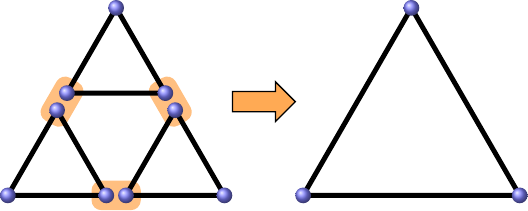}
 \caption{\label{fig:connect} 2D quantum repeater scheme based on three-party GHZ states. Short-distance GHZ states are connected to form a long-distance GHZ state with reduced fidelity, which is then re-purified to the initial fidelity via entanglement purification.}
\end{figure}

In the standard operational mode, mode I, a certain type of entangled state, e.g. a $m$-qubit GHZ state
\be
|GHZ_m\rangle = \tfrac{1}{\sqrt{2}}(|0\rangle^{\otimes m} + |1\rangle^{\otimes m}),
\ee
is produced at larger and larger distances, in such a way that the number of entangled parties (and type of state) is kept constant. This corresponds to a coarse graining of the lattice and the entanglement structure, where intermediate qubits are projected out (similar as in entanglement swapping). For $m=3$ qubit GHZ states, this is shown in Fig. \ref{fig:connect}. This process is performed similarly as in pairwise quantum communication using quantum repeaters. Elementary (multipartite) states are generated via direct transmission of qubits over noisy quantum channels. Several such short-distance entangled states are then connected such that the resulting state is the same as initially, but shared between parties at longer distance. The distance (in all directions) is thereby at least doubled. If the initial state or the local operations used for connection are not perfect, the fidelity of the resulting state is reduced. One can use MEP \cite{Du03,Du05,Kr06a,Du07,Glancy2006} to generate a state with the same fidelity as the initial elementary ones from several copies by means of local operations, thereby resulting in a situation as initially, however with entangled states of longer distance. This defines the 2D repeater cycle, which is applied in a concatenated way to achieve long-distant entanglement. As for the 1D repeaters, this approach leads to a polynomial scaling in the overall resources in the covered area and distance \cite{Br98}.

When using entanglement purification protocols with two-way classical communication, or a probabilistic connection procedure (see below), classical communication between the involved parties in a purification or connection step is required before states can be used at the next repeater level. This classical communication, together with gate times and preparation times of elementary states, determines the achievable rates. Notice that when using deterministic connection operations and deterministic entanglement purification with one-way classical communication, all steps of the protocol can be done simultaneously, and only a final correction operation at the end nodes (that can be done later) is required.

 %\subfloat[\centering]{\includegraphics[width=0.48\columnwidth]{connect_triangles_and_merge.pdf} \label{fig:connect}}
 %\hfill
 %\subfloat[\centering]{\includegraphics[width=0.48\columnwidth]{multilayers.pdf} \label{fig:illu_layers}}

 %\vspace{-3mm}
 %\begin{tabular}{m{0.08\columnwidth} m{0.92\columnwidth}}

 %{\hfill \small (c)} &   \subfloat{\includegraphics[width=0.9\columnwidth]{cluster.pdf} \label{fig:cluster} }
 %\end{tabular}
 %\caption{(a) 2D quantum repeater scheme based on three-party GHZ states. (b) On a triangular lattice, the GHZ states at different scales form a communication network where all parties can participate. (c) Repeater scheme based on 8-party 2D cluster states. Multipartite entangled states are connected via Bell-type measurements (light red), and some qubits are measured in the $y$ or $z$ basis.}
 %\vspace{-3mm}
%\end{figure}

\subsubsection{Flexible quantum communication network}
\begin{figure}[ht]
 \centering
 \includegraphics[width=\columnwidth]{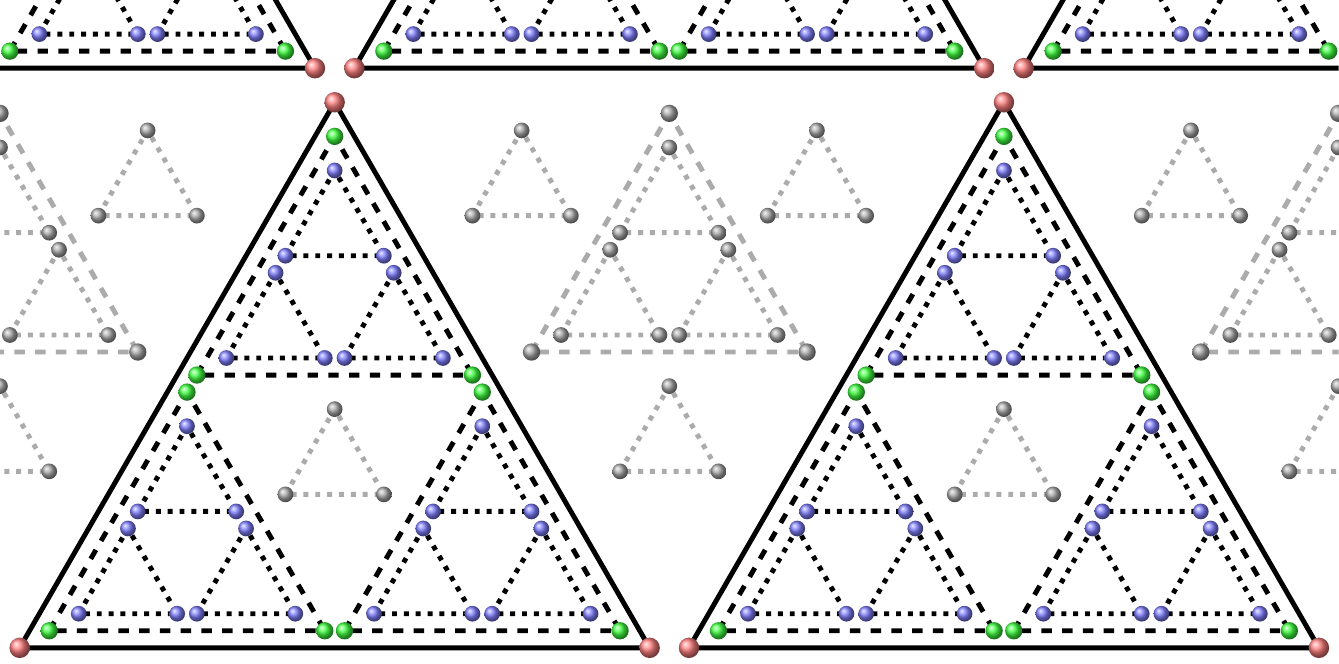}
 \caption{\label{fig:illu_layers} On a triangular lattice the GHZ states at different scales form a flexible communication network where all parties can participate.}
\end{figure}

A flexible quantum communication network where all partners are able to participate (and not just the outermost, far distant ones) can be achieved by using states from different scales, i.e. also the ones that are produced during earlier repeater cycles on short scales. These states can be combined such that GHZ states or pairs shared between {\it any} parties can be generated, see Fig. \ref{fig:illu_layers}. We remark that in such a flexible network, additional purification steps might be required, in particular when GHZ states of different scales are connected.

\subsubsection{2D cluster states on a rectangular lattice}
\begin{figure}[ht]
 \centering
 \includegraphics[width=\columnwidth]{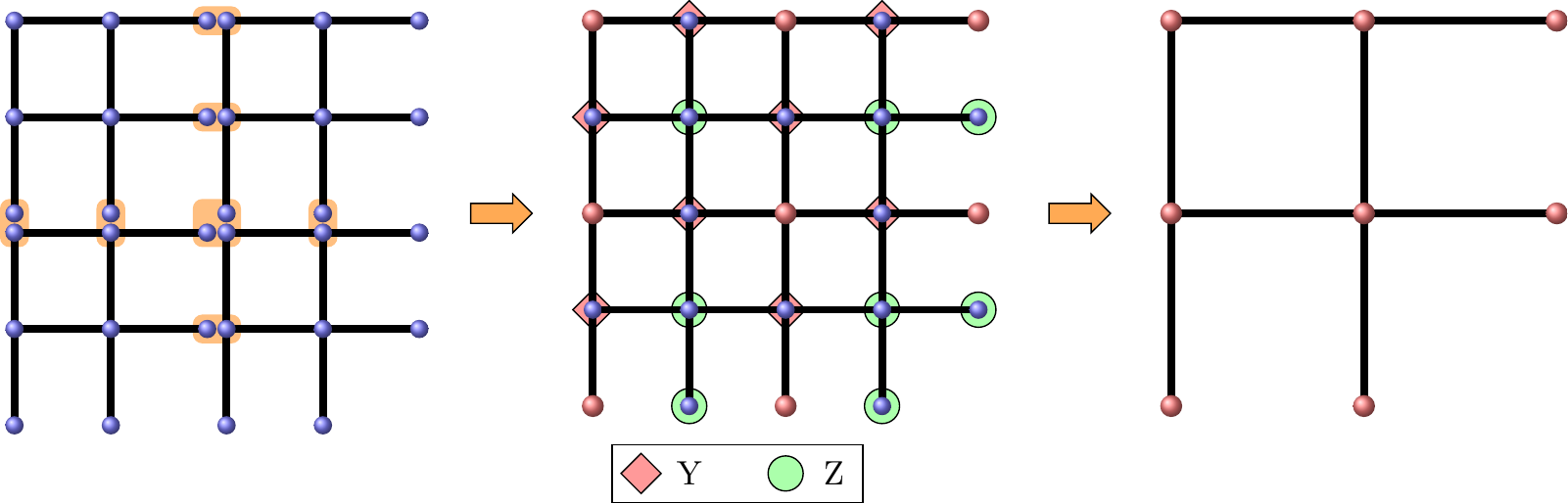}
 \caption{\label{fig:cluster} Repeater scheme based on 8-party 2D cluster states. Multipartite entangled states are connected via Bell-type measurements (light red), and some qubits are measured in the $y$ or $z$ basis.}
\end{figure}

The operational mode I is not restricted to the generation of GHZ states, but is also applicable to other entangled states such as graph states \cite{He06}
$|G_m\rangle=\prod_{(k,l)\in E}CZ^{(kl)}|+\rangle^{\otimes m}$,
where the edge set $E$ corresponds to the edges of a corresponding graph and determines the entanglement features of the state, and $CZ^{(kl)}=diag(1,1,1,-1)$ is the controlled-$Z$ operation acting on qubits $k,l$ (see appendix \ref{sec:graphstates}). An example for such a self-similar growing structure for a 2D-cluster type state on a rectangular lattice is shown in Fig. \ref{fig:cluster}.

\begin{figure}
\centering
 \includegraphics[width=\columnwidth]{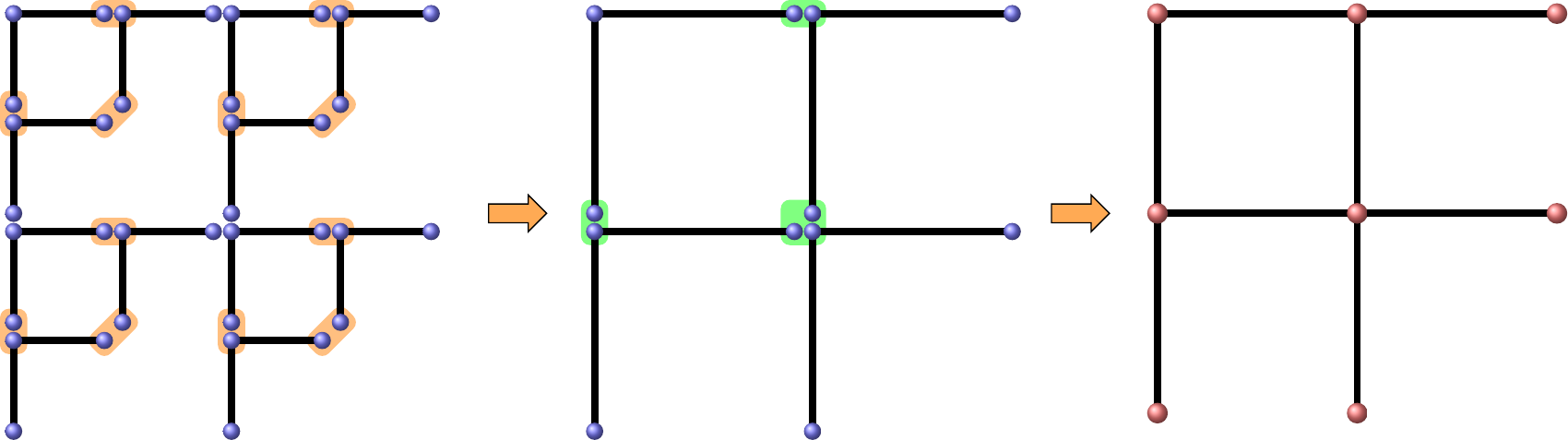}
 \caption{\label{fig:quad_ghz} The repeater scheme based on GHZ states implemented on a quadratic lattice can be used to construct coarse
	  grained 2D-cluster states. The L-shaped 3-qubit graph states are LU-equivalent to GHZ states.}
\end{figure}

The procedure to generate more and more coarse grained 2D-cluster states is conceptually very appealing because
the states at every repeater level have the same structure and one can observe the growth of a self-similar structure.
However, from figure \ref{fig:cluster}
it is apparent that many of the qubits are not involved in the protocol at all and are only
measured out to disentangle them. As an alternative scheme not relying on so many redundant qubits the repeater scheme for three-qubit GHZ states
can be used on a quadratic lattice and four GHZ states of an appropriate level can be combined to form the same coarse grained building
block (see figure \ref{fig:quad_ghz}).

\subsection{Alternative operational mode II}
Operational mode II corresponds to the growth of entangled states of similar type, but with increasing particle number. In that case, all particles at a specific site are merged into one. For the example of 3-qubit GHZ states on a triangular lattice, this means that starting from three $|GHZ_3\rangle$ states, one $|GHZ_6\rangle$ state shared between all nodes that are included in the larger triangle is generated (rather than a $|GHZ_3\rangle$ state shared between the nodes of the big triangle only). Similarly, the basic 2D-cluster structure is merged into a larger 2D cluster state with open links at the right and bottom (to connect it to neighboring structures), see Fig. \ref{fig:cluster} middle. Again, the fidelity is reduced and re-purification of the resulting states might be required. Notice, however, that now states are {\it not} the same as initially, and in fact MEP becomes less efficient for larger particle numbers. The threshold value, i.e. the tolerable error of local operations, becomes smaller for increasing system size for GHZ states \cite{DB04, He05}, essentially limiting the maximum size $m$. The threshold for 2D cluster states, in turn, is independent of system size \cite{DB04, He05}. This allows for the production of 2D cluster states of arbitrary size, which can e.g. be used as a resource for (distributed) measurement-based quantum computation \cite{Raussendorf2001,Briegel2009}.

%The required connection operations are described in Fig. \ref{fig:connect} and \cite{Sup}. Notice that measurement outcomes can also be used for error detection and the design of a probabilistic connection scheme.

\subsection{Connection of states to the next repeater level}
The required connection operations for operational mode I and II can be realized as follows. Two GHZ states $|GHZ_m\rangle \otimes |GHZ_n\rangle$ can be deterministically connected in such a way that (i) two qubits are merged into one, or (ii) both systems are projected out. In case (i) a projection $P_{S}=|00\rangle\langle 00| + |10\rangle \langle 11|$  or $P_S^\perp = |00\rangle\langle 01| + |10\rangle \langle 10|$ that acts on one qubit of the first, and one qubit of the second GHZ state is applied. The first qubit remains and the second is factored out, resulting in $|GHZ_{m+n-1}\rangle$  or $\eins^{\otimes m}\otimes \sigma_x^{\otimes n-1}|GHZ_{m+n-1}\rangle$ depending on the measurement outcome. The local Pauli operators can be corrected and a deterministic merging of two GHZ states is achieved. In case of (ii), one applies in addition a projection in the $X$ basis on the remaining qubit. (Equivalently, a Bell measurement, i.e. a measurement in the basis $\{|\Phi^{\pm}\rangle,|\Psi^\pm\rangle\}$ with $|\Phi^\pm\rangle = (|0\rangle\otimes|0\rangle \pm |1\rangle\otimes|1\rangle)/\sqrt{2}, |\Psi^\pm\rangle = (|0\rangle\otimes|1\rangle \pm |1\rangle\otimes|0\rangle)/\sqrt{2}$, can be directly applied to both qubits.) This leaves the remaining system in a $n+m-2$ particle GHZ state up to local Pauli operations that can be corrected. Notice that the results of the measurements can be used for error detection and the design of probabilistic connection schemes. For general graph states with an open link, i.e. a particle $A$ that is only connected to a single neighbor, a merging operation can be performed by first connecting $A$ via a $CZ$ operation to particle $B$ of the second graph state, and then measuring $A$ in the $Y$ basis \cite{He06}.

Merging all open links of the 2D cluster type state to the neighboring ones as illustrated in figure \ref{fig:cluster} leaves us with a larger 2D cluster state. The coarse-grained 2D cluster state can be obtained by additional $Z$ and $Y$ measurements. %Notice that the scheme we illustrated for 3-party GHZ states on a triangular lattice can also be applied to a rectangular lattice. In this case, 3-party GHZ states arranged on the rectangular lattice can be merged to form either a growing 2D cluster state, or a 3-party GHZ state of larger scale, where growth now occurs in $x$ and $y$ direction.

%\jwal{I added some sentences highlighting that this is for the particular scheme we use for the rest of the paper and tried to better explain what "the probabilistic scheme" is.Alternatively we could move it to the next section where we exclusively concern ourselves with that scheme (but it also does not really fit there.) }

The particular variant that connects three GHZ states to one GHZ state at the next repeater level will be analysed in detail in the sections below.
The protocol that does so deterministically can be summarized as follows

\begin{figure}[ht]
\centering
 \includegraphics[width=0.5\columnwidth]{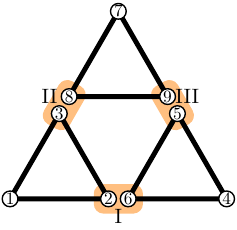}
 \caption{\label{fig:triangle_numbers} Repeater scheme connecting three GHZ states to one GHZ state on the next higher level.}
\end{figure}

\begin{itemize}
 \item Start with three copies of a (probably noisy) GHZ state.
 \item Perform Bell measurements on qubits (2,6), (3,8) and (5,9). (see figure \ref{fig:triangle_numbers})
 \item Depending on the outcomes of the Bell measurement perform correction operations as outlined in table \ref{tab:corr}.
\end{itemize}

Notice that only two of the Bell measurements are necessary to connect the three GHZ states. The seemingly redundant third
measurement does not only make the protocol symmetric but can actually be used to detect some specific errors. While it is not possible to correct the errors detected this way because the error syndromes are not unique,
it allows to discard the cases where an error is detected and therefore obtain better error thresholds.
However, this also means that the connection procedure only works probabilistically, and the whole procedure has to restart from the beginning if errors are detected.
The cases discarded are those in table \ref{tab:corr} with one or three $\ket{\Psi^\pm}$ outcomes.

Thus one may either use a deterministic connection procedure with slightly worse error thresholds,
or a probabilistic procedure where error thresholds are better, however, the rates may differ.
Unless explicitly stated all results in this paper use the probabilistic scheme.
Notice that as entanglement purification is already a probabilistic procedure, the performance and in particular the scaling of the overall 2D repeater scheme remains unchanged.

\begin{table}[ht]
 \begin{tabular}{ccc|c}
  Bell I & Bell II & Bell III & Correction \\
  \hline
  $\Phi^\pm$ & $\Phi^\pm$ & $\Phi^\pm$ & $\mathbbm{1}$ \\
  $\Phi^\pm$ & $\Phi^\pm$ & $\Psi^\pm$ &  - \\
  $\Phi^\pm$ & $\Psi^\pm$ & $\Phi^\pm$ &  - \\
  $\Psi^\pm$ & $\Phi^\pm$ & $\Phi^\pm$ &  - \\
  $\Psi^\pm$ & $\Psi^\pm$ & $\Phi^\pm$ &  $X^{(1)}$ \\
  $\Psi^\pm$ & $\Phi^\pm$ & $\Psi^\pm$ &  $X^{(4)}$ \\
  $\Phi^\pm$ & $\Psi^\pm$ & $\Psi^\pm$ &  $X^{(7)}$ \\
  $\Psi^\pm$ & $\Psi^\pm$ & $\Psi^\pm$ &  randomly one of $\{X^{(1)}, X^{(4)}, X^{(7)} \}$ \\
  \hline
 \end{tabular}
 \begin{tabular}{c|c}
  \# of $\Phi^-$ or $\Psi^-$ outcomes & Correction \\
  \hline
  0,2 & - \\
  1,3 & \pbox{20cm}{$Z$ on one of the remaining qubits \\ (does not matter which one)} \\
 \end{tabular}
\caption{\label{tab:corr} The correction operations for the repeater protocol connecting GHZ states (see figure \ref{fig:triangle_numbers}). }
\end{table}

\section{Analysis of 2D repeaters \label{sec:analysis} }
We now analyze the performance of the 2D repeater when taking noise and imperfections into account. %We consider two possible implementations of the scheme using (a) a gate-based approach and (b) a measurement-based approach \cite{Zw12,Zw13,Zw14,Zw15}. In (b), locally generated entangled resource states are used to perform the local connection operations and MEP via measurements only.
There are several relevant figures of merit for a repeater scheme. Here we concentrate on error thresholds for local operations and channels as well as reachable fidelities, as this provides information on whether such a scheme is suitable in principle. Another important quantity are achievable rates, which however depend strongly on the specific MEP and details of the implementation. We do not provide a full rate analysis for the general scheme here. For a concrete implementation with trapped ions, however, we also investigate distribution times and reachable distance with limited resources (see Sec. \ref{ions}).

We demonstrate that the usage of 3-party entangled states offers (in certain parameter regimes) an advantage over bipartite schemes w.r.t. error tolerance and achievable fidelity, but also for storage resources. This implies that there exist parameter regimes for channel noise and noisy local operations where a 2D approach allows one to generate GHZ states with a certain fidelity, while this is not possible with a 1D approach. Clearly, in this case also the achievable rates using the 2D approach are higher. In other regimes where both approaches are applicable, the achievable rates using a 1D approach might be higher, as multipartite recurrence type entanglement purification protocols are rather inefficient \cite{Du07}. The situation of direct state distribution (without a repeater scheme) using bipartite and multipartite strategies was investigated in \cite{Kr06} where a similar behavior was found.

\subsection{Error model}
Quantum channels are considered to be lossy and noisy, which prevents a direct transmission of quantum information over longer distances.
In addition, local operations at individual nodes of the network (parties) are considered to be noisy as well.
We model channel errors by a completely positive map (CPM) of the form
${\cal E}_q^{(a)} \rho=q\rho+\tfrac{1-q}{4}\sum_j \sigma_j^{(a)} \rho \sigma_j^{(a)}$,
with channel noise parameter $q$. We describe a noisy operation by ${\hat U}\prod_a{\cal E}_p^{(a)} \rho$, i.e. single-qubit local depolarizing noise (LDN) with error parameter $p$ on all involved particles, followed by the ideal operation described by the superoperator ${\hat U}$ with ${\hat U}\rho = U \rho U^\dagger$.

Clearly, in a physical realization errors may have different form, or channel losses may be dominant. However, this simple error model assuming depolarizing local noise covers the essential features and allows us to illustrate the effect of noise on the performance of such a 2D quantum communication network, similarly as done in \cite{Br98,Du07,Zw12,Zw15} for 1D repeaters. Notice that loss errors can in principle be mapped to depolarizing errors by replacing a lost qubit by a completely mixed state, but there are more efficient or practical ways to deal with loss, e.g. by just repeating the transmission as we consider in the trapped ion implementation below.

\subsection{Error thresholds}
In the following, we will concentrate on operational mode I. In order to analyze whether the repeater works despite imperfections in channels and operations, it is
useful to start by identifying noise thresholds that indicate up to which noise level states remain distillable.

\subsubsection{Error thresholds for entanglement purification}
%\jwal{come to think of it, calling the parameter in this section $p$ might be confusing as we use that only for the gate error in the rest of the paper. Suggestion: call it $q$ instead}
For noisy entangled pairs and GHZ states, necessary and sufficient conditions for distillability are known \cite{Du99, Du00, DB04, He05}.
In the case of local depolarizing noise with error parameter $q$ that acts on each of the particles, one finds a threshold of $q = 1/\sqrt{3}\approx 0.5774$ [$q\approx 0.5567$] of
local noise per particle for entangled pairs and 3-party GHZ states respectively \cite{Du99, Du00, DB04, He05} (see Appendix \ref{GHZdist}). For GHZ states, the threshold
value for $q$ increases with the size $m$ of the state, while for 2D cluster states the threshold for distillability is independent of $m$ \cite{DB04, He05}. A lower bound
on $q$ for distillability of 2D cluster states is given by $q=0.8281$ \cite{DB04,He05}.
Notice that, perhaps surprisingly, 3-party GHZ states are more robust against local noise than entangled pairs.
This implies that the error thresholds for MEP are more favorable than the ones for bipartite entanglement purification, and the use of 3-party states offers an advantage compared to the use of entangled pairs.

\subsubsection{Error thresholds for repeater cycle}
We now consider a repeater cycle, where the connection of three 3-party GHZ states is followed by appropriate MEP.
All involved operations are considered to be noisy, where we consider single qubit and two-qubit CNOT operations as elementary gates.
To obtain a threshold for the repeater cycle in the gate based model, a specific purification protocol has to be considered.
The threshold is determined by the amount of acceptable noise per gate, such that after connection of three elementary states followed by MEP,
the resulting state is still entangled and has at least the same fidelity as initially. Notice that we make use of the fact that the connection
of three GHZ states offers an intrinsic error detection capability, which allows one to obtain higher fidelities at the price that also the connection
procedure is non-deterministic. For the alternating MEP protocol \cite{Du03,Du05} we find $p_\mathrm{th} \approx 0.9581$. With more advanced
MEP schemes \cite{Reiter} (see appendix \ref{sec:adaptive}), this can be enhanced to $p_\mathrm{th} \approx 0.9490$. Then, the threshold for channel noise $q_\mathrm{min}$ at
the lowest level depends on $p$ and also the specific protocol used, where before the first connection an additional MEP is applied. While the direct multipartite approach
using these MEPs may not be optimal, it should be noted that for $p$ close to $1$ some of them already allow for a better $q_\mathrm{min}$ than is fundamentally possible
for a bipartite approach. Additional explanations and results are
provided in appendix \ref{MEPPReiter}.

\subsubsection{Basic repeater without entanglement purification}
\begin{figure}[ht]
 \centering
 \includegraphics[width=\columnwidth]{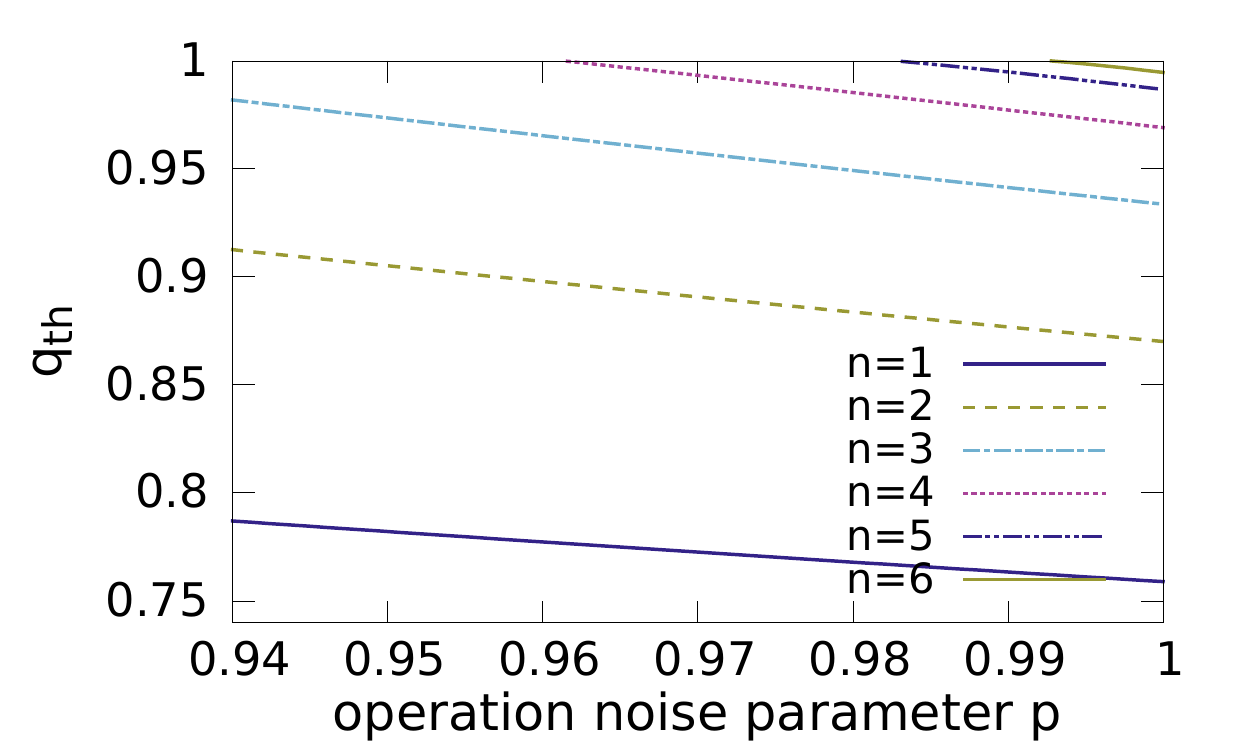}
 \caption{\label{fig:multi_levels} GHZ states are connected $n$-times in a concatenated way, i.e. $3^n$ states are connected to form a three-party GHZ state of distance $2^n$. The thresholds before state becomes disentangled is shown.}
\end{figure}

\begin{figure}[ht]
 \centering
 \includegraphics[width=\columnwidth]{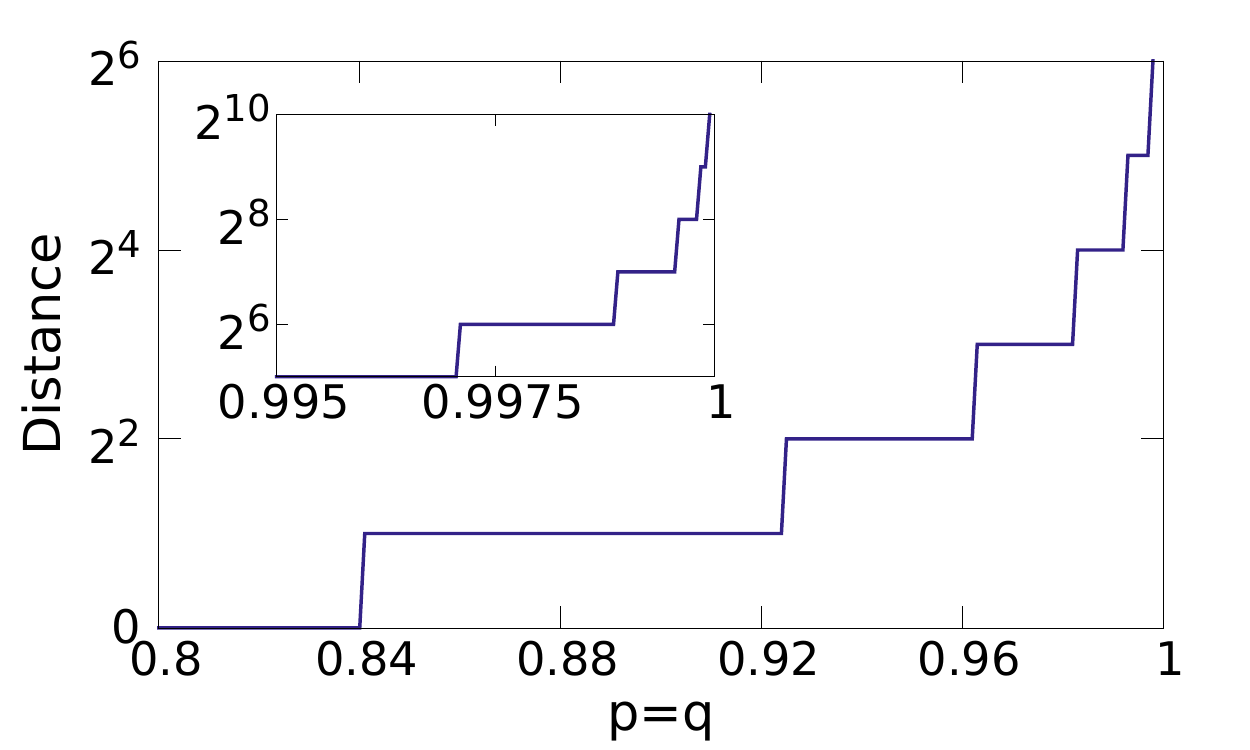}
 \caption{\label{fig:steps} GHZ states are connected $n$-times in a concatenated way, i.e. $3^n$ states are connected to form a three-party GHZ state of distance $2^n$. The maximum distance $2^n$ before state becomes disentangled is shown, where the given numbers correspond to multiples of elementary distance $L_0$ of the elementary GHZ states.}
\end{figure}

Another interesting quantity to look at is the number of iterations leading to the next repeater level can be performed before entanglement purification becomes necessary. This also provides thresholds for repeater schemes that operate without entanglement purification.
Figure \ref{fig:multi_levels} shows the thresholds for the local noise parameters $p$ and channel noise $q$ for different numbers of connection operations
such that the state remains distillable. The maximal reachable distance
is shown in figure \ref{fig:steps} for $p=q$.

%We finally remark that one can also consider a measurement-based implementation \cite{Zw12,Zw13,Zw14,Zw15} of the 2D repeater scheme using certain entangled states as resources. As shown in \cite{Sup}, such a scheme can tolerate up to 25\% noise per particle for resource states, which is even higher than the 23\% for a 1D implementation.

\section{Measurement-based implementation \label{sec:measurementbased} }
One may also consider a measurement-based implementation of entanglement purification and connection \cite{Zw12,Zw13,Zw14,Zw15}. In such an approach, task-specific entangled resource states are used to perform the connection and the purification procedure. Information processing takes place by coupling input particles via Bell-measurements to the particles of the (locally prepared) resource state. Since all the operations used are Clifford operations, the resource states are graph states, and consist only of input and output qubits \cite{Zw12,Zw13,Zw14,Zw15}.

The sole source of noise is given by imperfect resource states (and imperfect Bell measurements), which we model by depolarizing noise with parameter $p$ acting on each of the particles of the resource state as described above. For bipartite entanglement purification and 1D quantum repeaters \cite{Zw12}, it was shown that such an approach offers very high error thresholds, more than $13\%$ noise per particle for fault-tolerant quantum computation \cite{Zw14}, and more than $23\%$ noise per particle for bipartite entanglement distillation \cite{Zw13}. Noise can in fact be moved from resource states to input states under Bell measurements, which leaves us with perfect protocols applied to slightly noisier input states, and noise only acts on output particles \cite{Zw13,Zw15}.
Noisy resource states --that are used to implement the desired connection or purification operations-- are considered to be of the form $\prod_a{\cal E}_p^{(a)}|\psi\rangle\langle \psi|$, i.e. local depolarizing noise acting on each of the particles of the perfect resource state. This leads to an exponentially decreasing fidelity w.r.t. number of qubits, and incorporates that multi-qubit resource states are more difficult to prepare.

The error thresholds can be determined by considering the thresholds of the ideal purification protocol and taking local noise on the output particles into account as we show in detail below. Notice that the thresholds for MEP and the full quantum repeater are fact the same, and are protocol independent. This was shown in \cite{Zw15} for the 1D repeater, and the same argument holds in the 2D case. This follows from the fact that MEP and connection can be merged into a single resource state of minimal size that consists of only input and no output particles. As we have already seen above, the threshold for MEP for a 3-party GHZ state is as large as $p_{c} = pq = 0.5567$, leading to a threshold for the 2D quantum repeater based on tripartite GHZ states of $p_{\rm th} \leq \sqrt{p_c} \approx 0.7461$. That is, local noise of more than $25\%$ per particle is acceptable for resource states, which is even higher than for bipartite strategies.

\subsection{Noise in the measurement-based implementation}
%\jwal{Added this subsection heading as there was no transition between the paragraphs.}
In a measurement based approach any completely positive map $\mathcal{M}$ acting on $n$ qubits
can be probabilistically implemented using the resource state $\rho_\mathcal{M} = \mathbbm{1} \otimes \mathcal{M} \ketbra{\Phi^+}^{\otimes n}$
and utilizing Bell measurements to read in the input state. All the maps we use, including connection and multiparty entanglement purification, consist only of Clifford gates. This implies that they can be deterministically realized in such a measurement-based setup using resource states of minimal size, involving only input and output particles \cite{Zw12,Zw13,Zw14,Zw15}.
Noise that affects this resource state
naturally alters the effective map that is implemented. In the case of a local noise channel $\mathcal{E}$ acting
on each qubit of the resource state the analysis is straightforward because local noise can be shifted
freely between the two qubits on which a Bell measurement is performed \cite{Zw13}:
\begin{equation}
 \mathcal{P}_B^{1,2} \mathcal{E}^{(1)} \rho = \mathcal{P}_B^{1,2} \mathcal{E}^{(2)} \rho
\end{equation}
with the superoperator $\mathcal{P}_B$ describing the projection on a Bell state. Therefore the noise on the
qubits functioning as the read-in of the resource state can be transferred directly to the input state and it
is easily checked that the effective map the noisy resource state implements is given by $\mathcal{E}^{\otimes m} \mathcal{M} \mathcal{E}^{\otimes n}$
for a map $\mathcal{M}$ with $n$ input and $m$ output qubits.

Thus we can simply consider local noise channels being applied to all input qubits followed by perfect purification and connection operations and
finally noise on the output qubits that still remain after the procedure. This makes the analysis of error thresholds and performance of such a measurement-based scheme particularly simple.

\subsection{Finite purification steps in the measurement based scenario}
%In the main text we discuss thresholds for a measurement based implementation, which follows directly from the performance of ideal protocols as noise can be moved to input pairs.
In principle, it is possible to perform several purification rounds, and all repeater steps at all scales with a single (large) resource state at each node. This leads to the asymptotic error threshold of 25\% LDN per particle for a 3D repeater based on 3-party GHZ states announced above. However, here we assume that we only perform one repeater step at once with a particular resource state, and only a limited (small) number of purification steps. This is relevant for a small scale implementation with limited resources, as the required resource state are small.

\begin{figure}
\centering
 \includegraphics[width=\columnwidth]{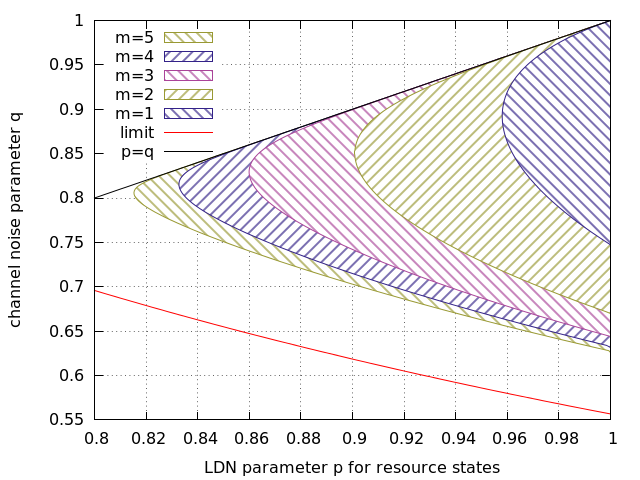}
 \caption{\label{fig:scenario_b} Parameter regions where implementing the repeater connection and $m$ purification steps using protocols P1 and P2 in a
	  measurement based way leads to an increase in fidelity.}
\end{figure}

We consider $m$ purification steps, where each step consists of the application of protocols P1 and P2. This is followed by the connection of the resulting GHZ states. Figure \ref{fig:scenario_b} shows the parameter region where this approach leads to an increase in fidelity after the first level, i.e. where a repeater cycle can be maintained.

\section{Comparison to architectures based on 1D repeaters \label{sec:1dcompare} }
We remark that one may also use an architecture based on a combination of 1D repeater schemes that allow one to establish Bell states between pairs of parties via bipartite entanglement purification and connection. Depending on the required task, this has to be done along a single connection line in the network (on-demand generation of pairwise entanglement), between several communication partners (generation of pairwise entangled states that are in a final step connected to form the desired multiparty state), or by establishing entangled pairs of various distance, and in different directions (network with pre-prepared bipartite states where all parties of the network can communicate).

There are many (intermediate) strategies how these tasks can be achieved. In order to compare the new 2D approach with 1D architectures, we consider the performance of both schemes under non-ideal conditions where state preparation, channels and gates are noisy, which we describe using the error model from above. This takes into account that the generation of elementary GHZ states is more difficult than the generation of entangled pairs. We use the recurrence entanglement-purification protocol of \cite{De96} in the 1D case, which is usually used in repeater schemes due to its large error thresholds and good performance \cite{Br98,Du07}.

We find that the direct generation of multi-party GHZ states as proposed here has clear advantages compared to strategies based on 1D repeaters with respect to several figures of merit. First, as shown in \cite{Kr06}, higher fidelities can be reached when the goal is to establish entangled states shared between three or more parties, e.g. 3-party GHZ states. Second, the required number of storage particles
per node is smaller, as we discuss below. Third, we have shown that MEP protocols and the whole repeater for the 3-party GHZ state admit higher error thresholds per particle than bipartite protocols.
Networks based on 1D quantum repeaters may still be more efficient in certain parameter regimes or for specific tasks, e.g. the preparation of long-distance bipartite entanglement or certain multipartite entangled states (see \cite{Kr06} for an analysis of direct creation of multipartite states using bipartite and multipartite strategies). While the optimal strategy for a given task may well be a combination of bipartite and multipartite strategies, we have shown that in certain parameter regimes a direct 2D approach outperforms 1D strategies.

\subsection{Storage requirements for multipartite and bipartite networks}
 In a full 2D triangular network, where 3-party GHZ states between all nodes are available as illustrated in Fig. \ref{fig:illu_layers}, the storage of final states requires three particles per node and coarse-graining level. In contrast, a system based on 1D repeaters on a triangular lattice requires either 4 or 6 particles, depending on whether one includes connection lines in two or all three directions. In both cases, this does not take into account that several copies might be needed to perform entanglement purification.

The multipartite approach offers an advantage concerning storage, mainly because storing one qubit is enough to have a connection
to multiple parties. The specific advantage however depend on which features one demands from the network. If one wants to consider a full triangular network for 3-party GHZ states - that is after
constructing states of any coarse-graining level the structure formed is again a triangular lattice - it is necessary to use all the states depicted in figure \ref{fig:illu_layers}.
That means all the holes in the Sierpinski triangle structure that appears when looking at different repeater levels have to be filled in with states up to the highest level possible.
For this particular setup a repeater station must be able to store 3 qubits per coarse-graining level it participates in. In contrast, with a bipartite strategy building up a full network
needs connections in all 6 directions at each node, which means 6 qubits per coarse-graining level have to be stored at a repeater station. Notice that this takes into account only the resulting states, while several copies are required to perform entanglement purification.

To guarantee that every party has access to the network in some form, the full scheme described above is not necessary. For example in the bipartite case it suffices to build a network in two
directions to reach every node in the network. However, if one requires the network to be structured in a way that each repeater station is only one connection away from a
repeater station of the next higher level, some connections in the third direction are required as well so the overall scaling does not change. Even when dropping this requirement
that still necessitates four qubits per coarse-graining level to be stored at repeater stations for a network based on 1D repeaters.
Even the not optimized multipartite network still scales better, so there is definitely a
storage advantage for the multipartite strategy. A switch to a quadratic lattice does not change this, although the third direction along the diagonals would probably not be used for the bipartite approach as the longer base distance for these connections would result in higher loss and error rates.

\section{Implementation using trapped ions \label{ions} }
There are intense efforts gearing up all over the world to build small--scale quantum networks and to connect multiple nodes \cite{Sa11}. So far, it has been shown on different platforms how two remote nodes can be entangled and the basic building blocks of a 1D quantum repeater have been experimentally demonstrated \cite{Chou:2005aa, Moehring:2007aa, Hofmann72, Hensen:2015aa}.

In principle, the 2D quantum repeater can be implemented in any system with a quantum light-matter interface and with the ability to perform quantum gates and measurements at each node. While several approaches are very promising, including atomic ensembles in microcells \cite{Borregaard2015} or NV centers in photonic crystal cavities \cite{Li2015,PhysRevA.72.052330}, we concentrate here on a setup with trapped ions, where all relevant building blocks have already been realized \cite{RevModPhys.82.1209, Northup:2014aa}.
While a scalable long-distance implementation requires MEP, a limited approach based on the generation and connection of elementary multipartite states without entanglement purification still allows to obtain networks over considerable distances. We analyze such a quantum network below, which shows concretely how an ion-based 1D repeater \cite{PhysRevA.79.042340} can be extended to 2D networks.

\subsection{Distributing a GHZ state with the 2D quantum repeater}
\begin{figure}[htb]
%\label{rates}
\centering
\includegraphics[scale=0.3]{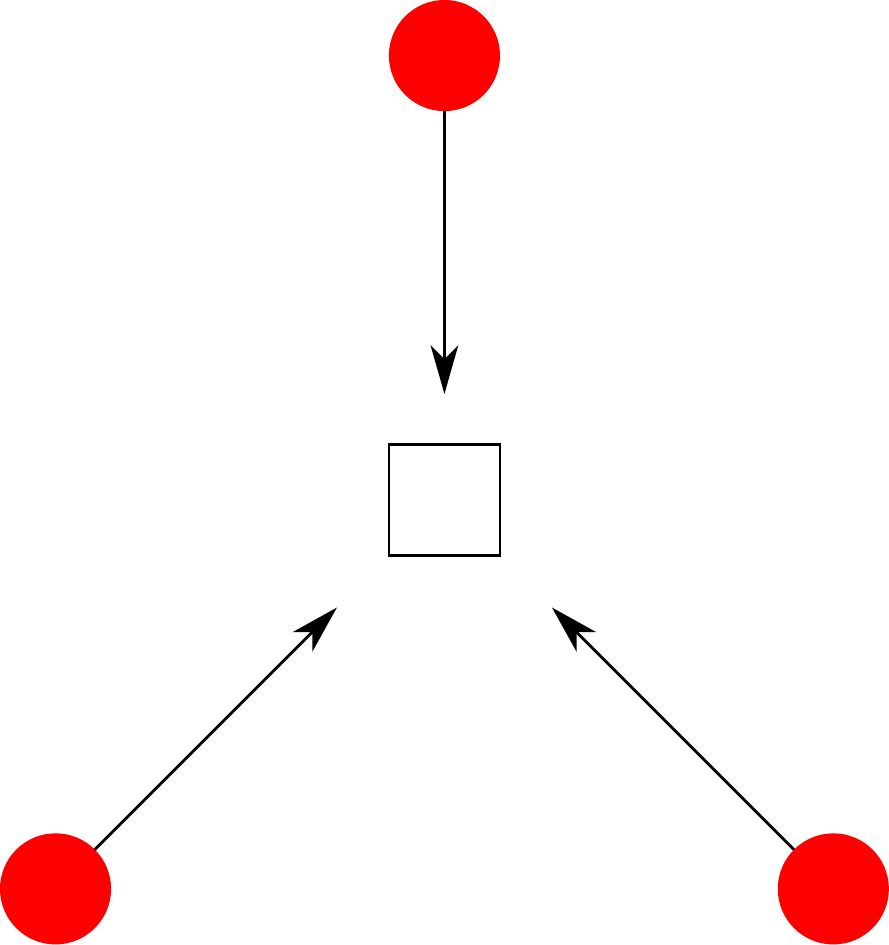}
\caption{Illustration of creating a three-qubit GHZ state. The box denotes the linear optics setup and photon detectors, the red dots represent the ions in the cavities.}
\label{GHZ3}
\end{figure}

We concentrate on a scheme based on the generation of 3-party GHZ states using ion-photon entanglement and a suitable linear optics setup for projecting the photons \cite{Wang2009}. An illustration is provided in Fig. \ref{GHZ3}. These elementary GHZ states are connected via swap operations. The swap operation is simply a (deterministic) Bell measurement, which can be implemented with an entangling gate and single-qubit measurements. As a simplified repeater scheme without entanglement purification is considered, the number of states that can be connected is limited.

%Strategies to establish other graph states in such a set-up are discussed below.

%Here, the elementary building block, a three-qubit GHZ state, is created according to the scheme proposed in \cite{Wang2009}.

We would like to mention that one can prepare arbitrary graph states, e.g the 8-qubit cluster state, by creating a Bell pair for each edge in the graph and subsequent local operations/projections at each node.

\subsection{Parameters}
We assume the following parameters: ion-photon entanglement with 99.5\% fidelity, single- (two-) qubit operations with 0.1 (0.5) LDN \cite{Stute2012,Schindler2013}, 90\% photon detector efficiency \cite{detector1,detector2}, 90\% probability of ion emitting single photon and successful frequency conversion to telecom wavelength \cite{Stute2012,Zaske2012,Fernandez13}, and standard telecom
fibers with attenuation length of 22 km. Notice that all these parameters have already been achieved in experiments, except the probability of emitting a photon which is however expected to be reachable with
present-day set-ups \cite{Northup:2014aa}.
For the comparison of the 1D and 2D approach, we remark that this error model takes into account that the generation of GHZ states is more difficult than the generation of entangled pairs.
\subsection{Fidelities}
%\jwal{removed the duplicate information and one of the plots showed redundant information}
%The resulting fidelities for 4, 8 and 16 links are given by $0.9096; 0.7803; 0.5877$. The distribution times are shown in Fig. \ref{figure_times}. This allows one to distribute a three-qubit GHZ state with a fidelity large enough to violate a Bell inequality \cite{Mer90,Bel93,Lanyon2014}, over a distance of more than 1000 km in less than one second.
%

% \subsubsection{Comparison of 1D and 2D strategy}
We compare our intrinsic 2D strategy with a 1D approach. In this setting one prepares Bell pairs between nodes A and B, and between nodes A and C using a 1D quantum repeater. The GHZ state between nodes A, B, and C can be created with an additional ancilla qubit in state $\ket{+}$ at node A and CNOT gates between this qubit, which serves as control qubit, and the two qubits, which are part of the shared Bell pairs, followed by a measurement of the target qubits in the computational basis.

Surprisingly, the 2D approach allows one to obtain slightly higher fidelities for the resulting long-distance GHZ state.
The numbers for various numbers of links are summarized in table \ref{compfid}.

\begin{table}[h]
%\label{tablefid}
\vspace{0.3cm}
\centering
\begin{tabular}{| c |c | c | c | }
\hline
& 4 links & 8 links & 16 links \\ \hline
1D & $86.44\%$ & $74.85\%$ & $57.18\%$ \\  \hline
2D & $90.96\%$ & $78.03\%$ & $58.77\%$ \\
\hline
\end{tabular}
\caption{\label{compfid} Fidelities for several numbers of links using the 1D and the 2D approach.}
\end{table}

Notice that this holds for negligible memory errors. For ion traps the dominant error source is collective dephasing \cite{Monz2011}, so that one needs to either assume that the coherence time is large compared to the distribution time or encode quantum information into a decoherence-free-subspace \cite{Zanardi1997,Lidar1998} (see also \cite{DFS}). We have performed an analysis of the influence of memory errors and the usage of a decoherence-free-subspace encoding in \cite{DFS} for a 1D repeater scheme. The techniques are also applicable in the 2D approach presented here.

Note also that an all--optical implementation, generalizing the work of \cite{Azuma2015}, is conceivable, provided one uses a one-way entanglement purification protocol.

%-----------

%\subsection{Implementation with trapped ions}
%In the following we provide more details on the implementation of 2D quantum repeaters (without entanglement purification) with trapped ions. In particular we  discuss the derivation of the fidelity and distribution time and also compare the 2D quantum repeater to an approach based on 1D quantum repeaters.

%\paragraph{Comparison. ---}
%We find that the 2D approach leads to higher fidelities than the 1D approach.

\subsection{Distribution times}

The derivation of the distribution time of the GHZ states is analogous to \cite{DFS}. We use the parameters listed above.
% The fiber attenuation length is given by $L_{att}$ = 22 km, which leads to a transmission probability $\eta_t=\text{exp}(-L/L_{att})$ for a fiber of length $L$. The total probability of an ion emitting a single photon and frequency conversion is given by $p_{ion}=0.9$, and for the detector efficiency we assume $\eta_{d}=0.9$. The speed of light in the fiber is $c=2\cdot10^8$m/s. We also assume ion-photon entanglement with 99.5\% fidelity and single- (two-) qubit operations with 0.1 (0.5) local depolarizing noise.

% Notice that for the comparison with the 1D scheme, our error model takes into account that the generation of GHZ states is more difficult than the generation of entangled pairs. We have a smaller success probability (as two photon emissions and transmission need to succeed), and also a smaller fidelity.

\subsubsection{Distribution time using 2D quantum repeater}
\begin{figure}[ht]
  \centering
  \includegraphics[width=\columnwidth]{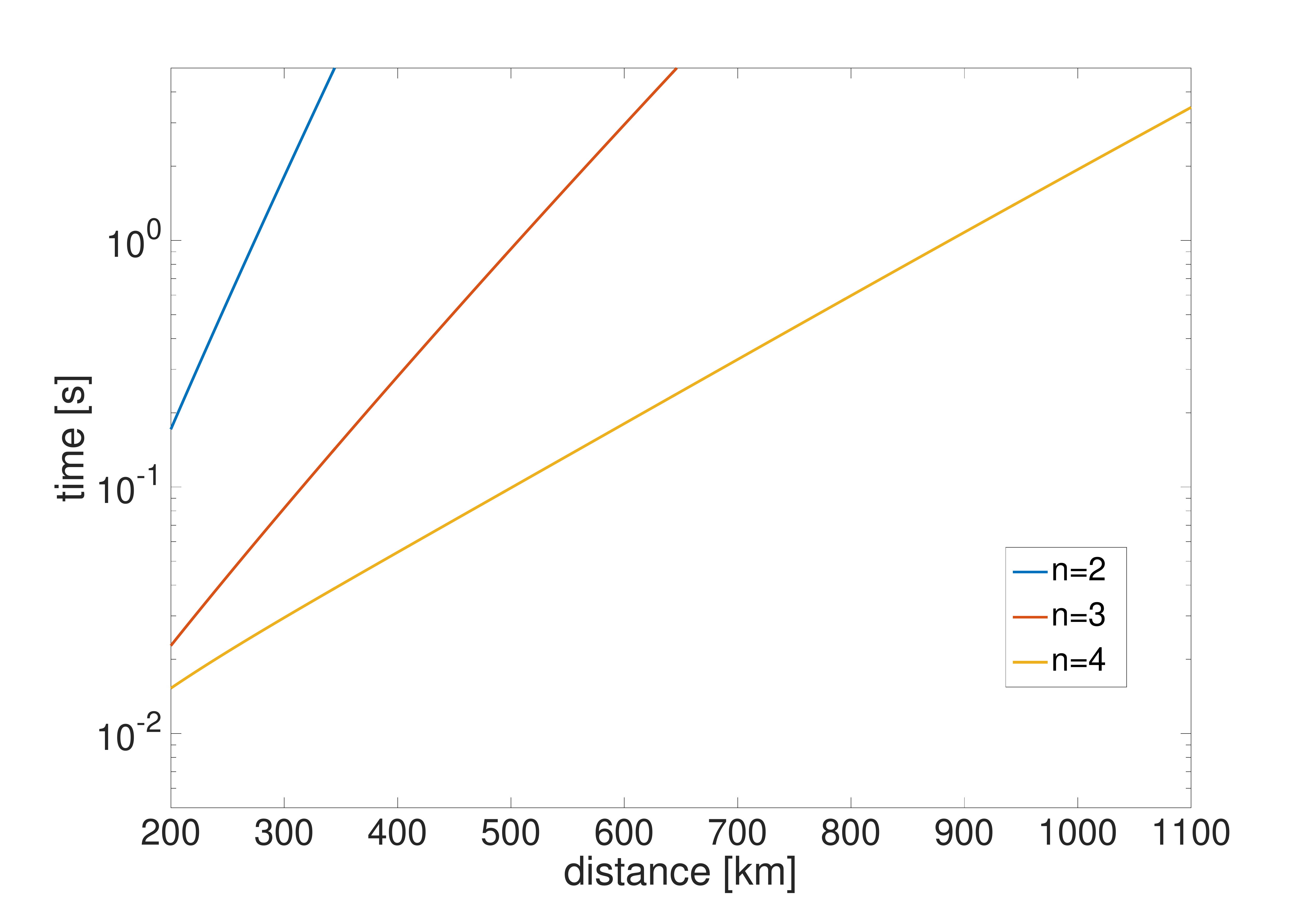}
  \caption{ \label{figure_times} Distribution time using an ion-trap based implementation as a function of the distance.}
 \end{figure}

In this approach the photons are sent from the corners of the triangle to the center, where the linear optics elements and photon detectors \cite{Wang2009} are placed (for concreteness we assume an equilateral triangle). The distance to the center is then given by $\tfrac{L_0}{\sqrt{3}}$, where $L_0$ is the distance between the nodes. The probability of creating the GHZ state between three elementary nodes is then given by
\be
P_{elem}=\frac{1}{4}p_{ion}^3 \eta_d^3\eta_t^3,
\ee
with $\eta_t=\text{exp}(-L_0/(\sqrt{3}L_{att}))$. The time for distributing the GHZ state is then given by (see also \cite{DFS})
\be
T=\frac{L_0}{c}\sum_{i=1}^{3^{n}}\frac{1}{1 - (1-P_{elem})^i},
\ee
for a repeater with $2^n$ links. The resulting distribution times are shown in Fig. \ref{figure_times}.

\subsubsection{Distribution time using 1D quantum repeater}
Here the probability of establishing an elementary Bell pair is given by
\be
P_{elem}=\frac{1}{2}p_{ion}^2 \eta_d^2\eta_t,
\ee
with $\eta_t=\text{exp}(-L_0/L_{att})$. The time for distributing the GHZ state is then given by (see also \cite{DFS})
\be
T=\frac{1}{p_{\rm suc}}\frac{L_0}{c}\sum_{i=1}^{2\cdot2^{n}}\frac{1}{1 - (1-P_{elem})^i},
\ee
for a repeater with $2^n$ links, where $p_{suc}$ is the total success probability for appropriate Bell measurement outcomes in the connection processes.

\subsubsection{Comparison of 1D and 2D strategy}
\begin{figure}[htb]
%\label{rates}
\centering
\includegraphics[width=\columnwidth]{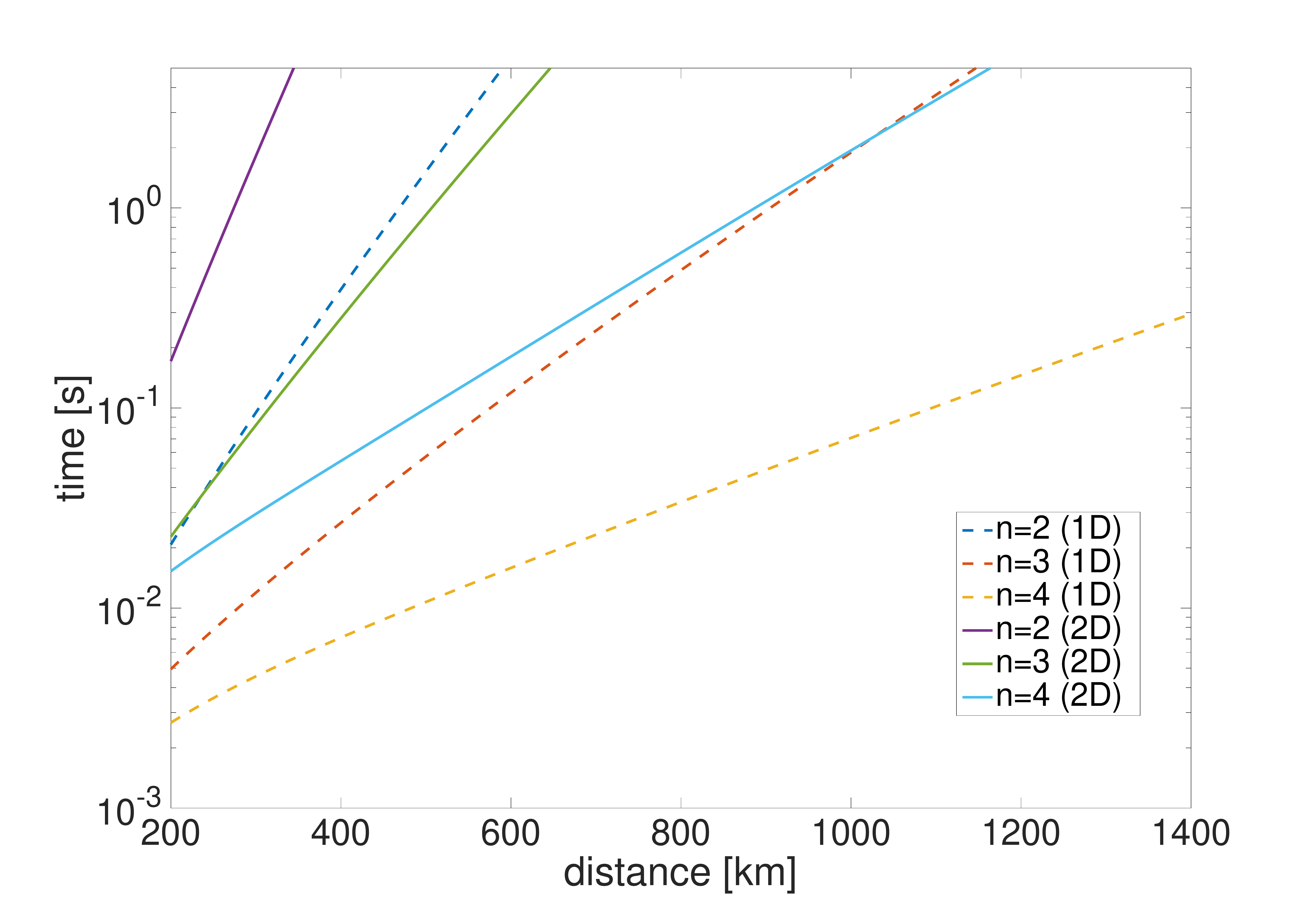}
\caption{Distribution time for a three-qubit GHZ state using the 1D and 2D quantum repeater as a function of distance.}
%\jwal{axes labels and line width could be bigger} }
\label{comptime}
\end{figure}

The distribution times for 4, 8 and 16 links are plotted in Fig. \ref{comptime}, both for the 1D and the 2D approach. The times for the 1D approach are clearly smaller, and it should also be noted that the 2D approach requires more resources (total number of nodes and ions). However with both approaches one can distribute a GHZ state with fidelity that is large enough to violate a Bell inequality \cite{Mer90,Bel93,Lanyon2014} over a distance of 1000 km in about one second.

\section{Summary and Outlook \label{sec:summary}}
We propose a 2D repeater architecture based on self-similar growing structures of multipartite entangled states. The favorable scaling and the high error thresholds of 1D repeater architectures do not only carry over to the 2D approach, 2D repeaters provide in fact higher error thresholds while at the same time, the required resources for storage are reduced. Furthermore, the design of quantum repeaters is not restricted to GHZ or 2D cluster states on 2D lattices. In fact, different lattice structures (including 3D arrays) and target graph states are conceivable.
Notice that MEP protocols for all graph states exist \cite{Du03,Du05,Kr06a,Du07}, such that the procedure of connecting and purifying states at different scales or growing size is universally applicable.

The proposed approach is intrinsically two-dimensional and seems thus ideally suited for real world quantum networks. In particular, such networks offer a high degree of flexibility, with potential two-party and multi-party applications. As outlined above, various platforms offer themselves for an implementation in the near future, opening the way towards real-world application of quantum technology in large-scale quantum networks.

{\em Acknowledgements} This work was supported by the Austrian Science Fund (FWF): P24273-N16, P28000-N27, by the Swiss National Science Foundation (SNSF) through Grant number PP00P2-150579, the Army Research Laboratory Center for Distributed Quantum Information via the project SciNet and the EU via the integrated project SIQS.

\bibliographystyle{apsrev4-1}
%\bibliography{2Drepeater}
%

\newpage
\section*{}
%\newpage

\appendix
%\section*{Supplementary material}

%In the supplementary material we discuss some details of the repeater scheme such as the distillability criterion for GHZ states and the exact formulation of the connection operation. We also briefly review entanglement purification schemes for multipartite graph states and present additional results for noise thresholds when using specific protocols. Furthermore, we discuss a measurement based implementation restricted to small resource states and provide an alternative way to generate coarse grained 2D-cluster states based on the repeater scheme for GHZ states. We also elaborate on the storage advantage a multipartite repeater network offers, and provide additional details on a possible implementation with trapped ions.

\section{Distillability of GHZ states}
\label{GHZdist}
For GHZ states there exists a standard form (in which all considered states are automatically due to the error models used):
\bea
 \rho&=&\lambda_{\bm{0}}^+ \ketbra{\Psi^+_{\bm{0}} } + \lambda_{\bm{0}}^- \ketbra{\Psi^-_{\bm{0}} } \\ \nonumber&&+ \sum_{ \bm{k} \neq \bm{0} } \lambda_{\bm{k}} \left( \ketbra{\Psi^+_{\bm{k}} } + \ketbra{\Psi^-_{\bm{k}} } \right)
\eea
 with $\ket{\Psi^+_{\bm{0}} } = \ket{GHZ_n}$, $k=(k_2,k_3,\dots,k_n)$ and $\ket{\Psi^+_{\bm{k}} } = \prod_i \left(\sigma_x^{(i)} \right)^{k_i} \ket{\Psi^+_{\bm{0}} }$ as well as
 $\ket{\Psi^-_{\bm{k}} } = \sigma_z^{(1)} \ket{\Psi^+_{\bm{k}} }$.
This state is distillable to a $n$-partite GHZ state iff \cite{Du99, Du00}:
\begin{equation}
 \lambda_{\bm{0}}^+  - \lambda_{\bm{0}}^- > 2 \lambda_{\bm{k}} \quad \forall \bm{k} \neq \bm{0}
 \label{eqn:distcrit}
\end{equation}

The distillation procedure proposed in \cite{Du99,Du00} that allows to distill all states that fulfill these criteria
has an initial multiparty step where
$P=|0\rangle\langle 0|^{\otimes N} + |1\rangle\langle 1|^{\otimes N}$ is applied on each
party of $N$ copies of the 3-qubit GHZ state, but then involves the purification of
two-qubit pairs that are recombined. The distabillity of these qubit-pairs
(i.e. fidelity greater than 0.5) is what sets the limit of the whole procedure.
An alternative approach, however, is to use multipartite entanglement purification \cite{Du03,Du05,Du07,Reiter} directly.
These purification schemes come very close to the theoretical limit and two of these protocols already clearly surpass the bipartite threshold (see figure \ref{fig:cycle}).

Using the criterion \eqref{eqn:distcrit} for a 3-qubit GHZ state that has been
effected by local white noise with error parameter $p$ on every qubit the
threshold turns out to be $p \approx 0.5567$, which is even better than the bipartite threshold of
$p=\frac{1}{\sqrt{3}} \approx 0.5774$.

\section{Graph states}
\label{sec:graphstates}

The graph state $\ket{G}$ corresponding to the graph G with vertices $V = \left\{ 1, 2, \dots , N \right\}$ and edges
$E \subseteq [V]^2$, where $[V]^2$ is the set of subsets of $V$ containing $2$ elements, is given by:
 \begin{equation}
  \ket{G} = \prod_{ \left\{a,b \right\} \in E } CZ^{ab} \ket{+}^{\otimes N}
 \end{equation}
 where $\ket{+}$ is the eigenstate of $\sigma_x$ with eigenvalue $+1$ and
 $CZ^{ab}=\ketbra{0} \otimes \mathbbm{1} + \ketbra{1}\otimes \sigma_z$
 is the controlled-$\sigma_z$ gate acting on qubits $a$ and $b$.

 The orthogonal \textit{graph state basis} is defined by:
  \begin{equation}
  \ket{\bm{\mu}}_G = \prod_{j \in V} \left( \sigma_z^j \right)^{\mu_j} \ket{G}
 \end{equation}
 with $\bm{\mu}=(\mu_1, \mu_2, \dots, \mu_N) \in \{0,1\}^N$. The subscript $G$
 acts as a reminder that these states are defined with respect to a particular graph.

\section{Adaptive MEP protocols}
\label{sec:adaptive}

In \cite{Du03,Du05} an entanglement purification protocol is introduced for two-colorable graph states.
This protocol is relevant in the context of the repeater scheme because the GHZ state is LU-equivalent
to a two-colorable graph state.
A density operator diagonal in the graph state basis is considered:
\begin{equation}
 \rho = \sum_{\bm{\mu}_A,\bm{\mu}_B} \lambda_{\bm{\mu}_A,\bm{\mu}_B} \ket{\bm{\mu}_A,\bm{\mu}_B}_G \bra{\bm{\mu}_A,\bm{\mu}_B}
\end{equation}
with the binary vector index $\bm{\mu}=(\mu_1,\dots,\mu_n)$ split in two parts $\bm{\mu}_A$ and $\bm{\mu}_B$ to emphasize the two sets of
qubits corresponding to different colors. The graph state $\ket{\bm{0},\bm{0}}_G$ is the desired state in this case and therefore
the fidelity is given by $\lambda_{\bm{0},\bm{0}}$.
The protocol consists of two subprotocols P1 and P2 that are applied in an alternating way. They consist of CNOT operations applied on two copies
of the graph state on every party followed by local measurements on the second copy. The measurement outcomes indicate whether the purification step was successful.
In short, the effect of P1 is
to amplify coefficients with $\bm{\mu}_A = \bm{0}$ while P2 amplifies the coefficients with $\bm{\mu}_B = \bm{0}$.

However, \cite{Reiter} shows that the alternating sequence is not ideal for all states and analyzes two adaptive schemes:

In the maximum local fidelity (MLF) adaptive scheme (which was already mentioned in \cite{Du05}) one simply uses the protocol, P1 or P2, that leads to a
higher fidelity.
The premise of the lambda weight (LW) adaptive scheme is the observation P1 and P2 increase different coefficients. In this scheme the choice between
P1 and P2 is made by comparing the sum of the coefficients associated with each protocol, that is applying P1 if
\begin{equation}
 \sum_{\bm{\mu}_A} \lambda_{\bm{\mu}_A,\bm{0}} > \sum_{\bm{\mu}_B} \lambda_{\bm{0},\bm{\mu}_B}
\end{equation}
and P2 otherwise.
While these adaptive protocols can increase the purification regime especially in the presence of noise, both
protocols cannot be directly applied in an experiment since both require information about the state that is not easily obtained.
However, if the initial states are known, computer simulations ahead of the experiment can find out which particular sequence of subprotocols
the schemes would suggest for that particular setting.

We compare the performance of the different MEP protocols in the context of the 2D repeater below.

\section{Repeater thresholds for particular purification schemes}
\label{MEPPReiter}
\begin{figure}[ht]
\centering
\includegraphics[width=\columnwidth]{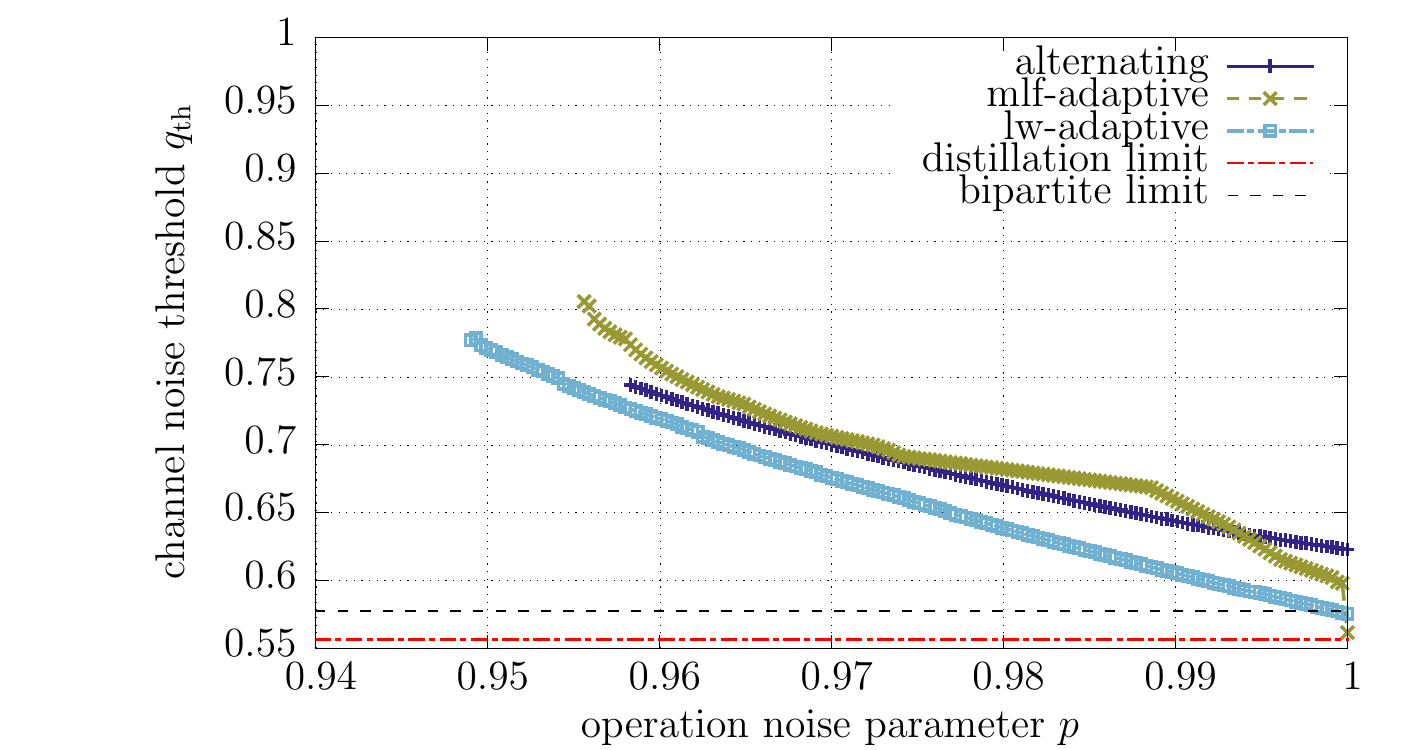}
 \caption{\label{fig:cycle} Thresholds for the repeater cycle when using different purification protocols.
	  The cut-off at low $p$ indicates the threshold $p_\mathrm{th}$ for that particular purification scheme.}
\end{figure}

\begin{figure}[ht]
\centering
\includegraphics[width=\columnwidth]{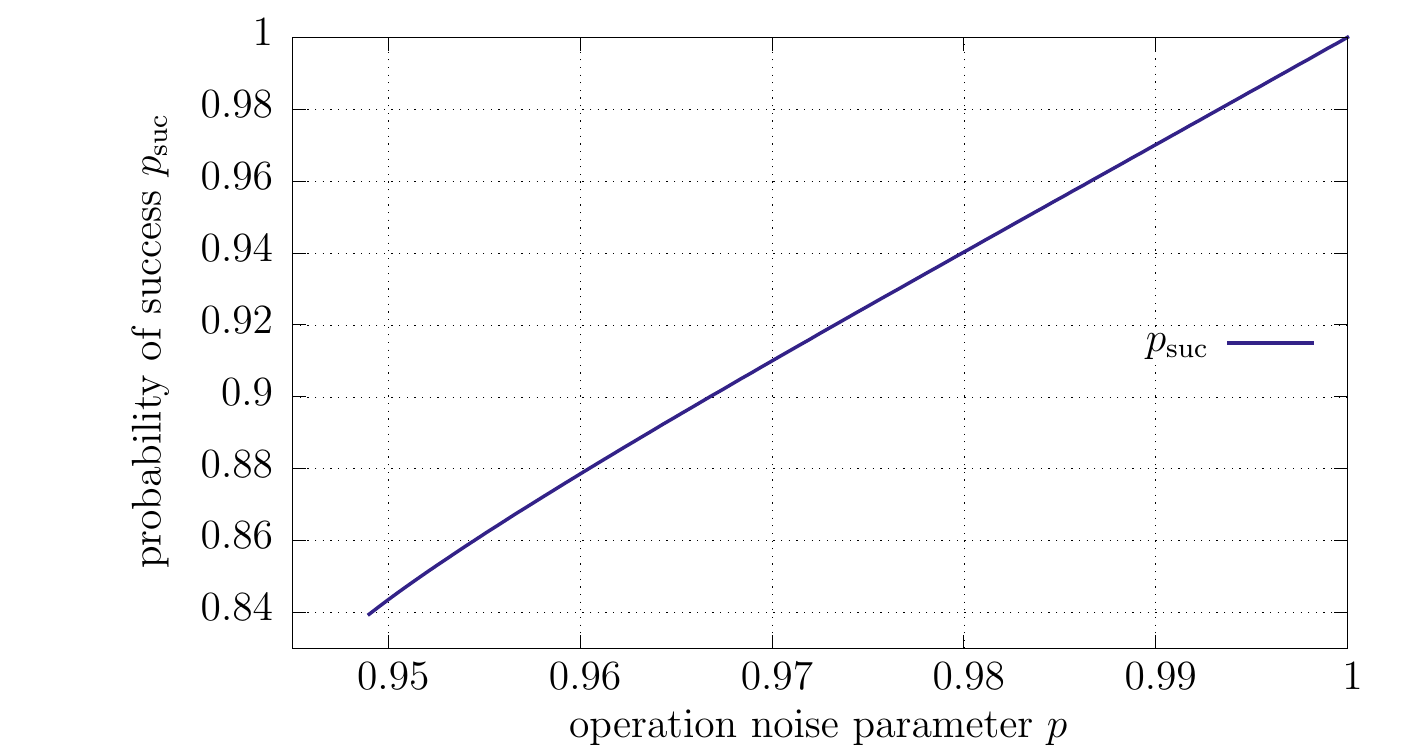}
\caption{\label{fig:p_suc_instant} Probability of success of the probabilistic connection operation in the repeater scheme using resulting states of MEP after multiple purification rounds (fixed point) with gate error parameter $p$ }
\end{figure}

We analyzed the repeater thresholds for specific MEP schemes in the gate based scenario.
The error parameter $p$ for the Bell measurements and the CNOT operations used in the MEP protocol are considered to be the same.
The threshold value $p_\mathrm{th}$ is the lowest $p$ for which the repeater cycle can be maintained.
While the purification
protocols themselves work for smaller $p$ the states they output at their fixed point can no longer be purified by the same protocol after the next repeater step
if $p<p_\mathrm{th}$.
For the alternating scheme introduced in \cite{Du03} we find $p_\mathrm{th} \approx 0.9581$. With the adaptive variants of \cite{Reiter}
the this threshold can be improved. The MLF-adapative scheme produces $p_\mathrm{th} \approx 0.9554$ and the LW-adaptive scheme $p_\mathrm{th} \approx 0.9490$.
Then, the threshold for the local noise $q_\mathrm{min}$ at the lowest level depends on $p$ and also the specific protocol used (see figure \ref{fig:cycle}).
Note that the purification regimes are cut off towards low $p$ at the value $p_\mathrm{th}$ corresponding to that particular scheme.
There might, however, be a multipartite entanglement purification scheme that performs better than these.
While the adaptive schemes come close to the distillation limit for $p=1$ they do not quite reach it.
The success probability of the probabilistic connection operation used at each repeater level (figure \ref{fig:p_suc_instant}) is the same for the three protocols.
As entanglement purification is already a probabilistic procedure,
the scaling behaviour and principle performance of the repeater are not effected.

In the main part we also discussed a setting where $n$ connection steps are performed, i.e. $3^n$ elementary GHZ states
are connected in order to form a GHZ state at distance $2^n$.
In table \ref{tab:relay_values} some exemplary values for the noise threshold $q_\mathrm{th}$ such that the resulting
state is still distillable are provided in addition to figure \ref{fig:multi_levels}.

\begin{table}
 \begin{tabular}{c|c|c}
  $n$ & $p=0.99$ & $p=0.995$ \\
  \hline
  $1$ & $0.7635$ & $0.7612$ \\
  $2$ & $0.8769$ & $0.8735$ \\
  $3$ & $0.9415$ & $0.9376$ \\
  $4$ & $0.9773$ & $0.9733$ \\
  $5$ & $0.9950$ & $0.9910$ \\
  $6$ & - & $0.9987$
 \end{tabular}
 \caption{\label{tab:relay_values} Threshold values $q_\mathrm{th}$ for the channel noise at the lowest level in a 2D repeater setting such that the
          resulting state after $n$ connection steps is still distillable.}
\end{table}

\begin{figure}[ht]
\centering
 \includegraphics[width=\columnwidth]{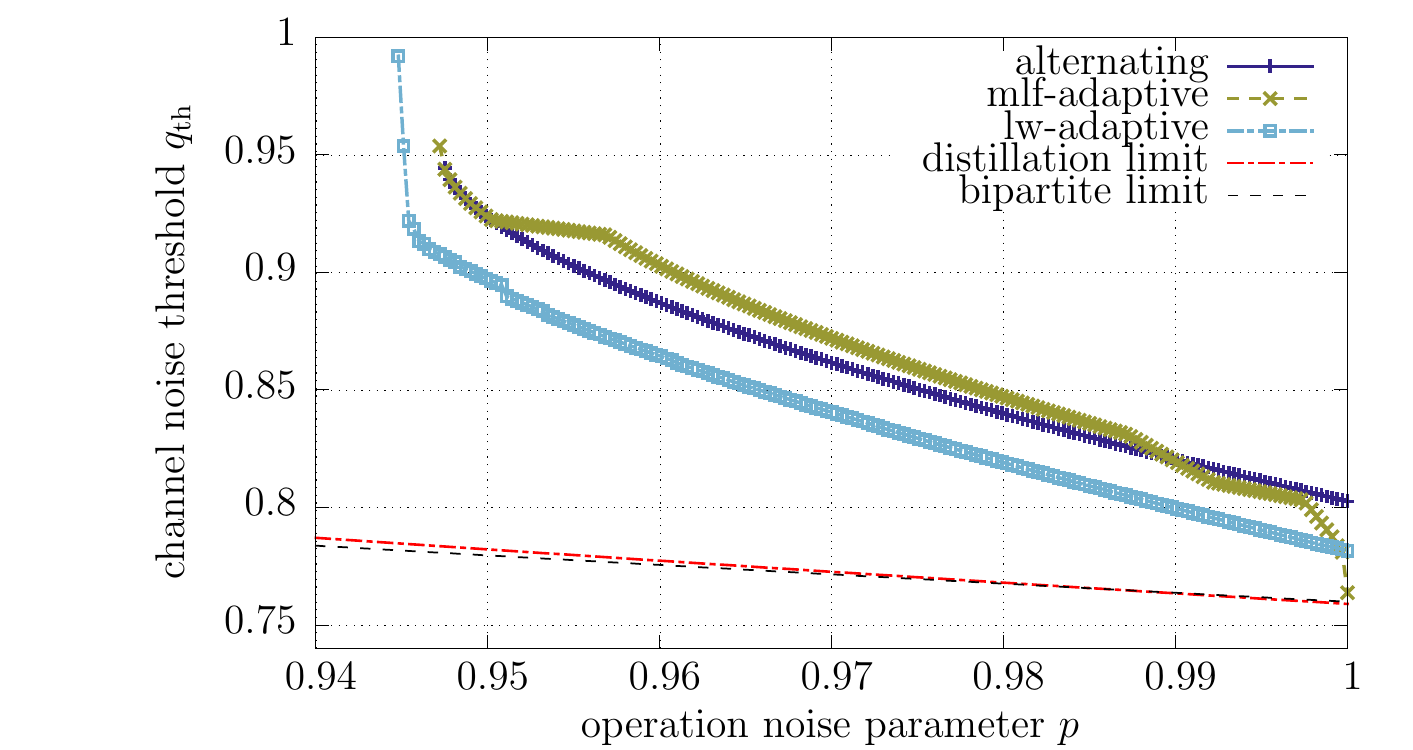}
 \caption{\label{fig:scenario_a} Thresholds for different entanglement purification protocols, indicating
	  that the state can still be purified after one initial connection procedure.}
\end{figure}

Furthermore, we also analyse the performance of different MEP protocols. We consider a situation where two states are first connected, and then purified using the specific MEP.
The analysis the case of performing one such operation is shown in figure \ref{fig:scenario_a}. Note however, this does not take into
account whether the resulting state after the purification is suitable for further repeater levels. The probability of success at each such step is shown in
table \ref{tab:p_suc_a} for some example values of $p=q$.

\begin{table}[ht]
 \begin{tabular}{c|c|c|c}
  $n$ & $p=q=0.98$ & $p=q=0.99$ & $p=q=0.995$ \\
  \hline
  $1$ & $89.24\%$  & $94.32\%$  & $97.08\%$   \\
  $2$ & $88.48\%$  & $94.12\%$  & $97.02\%$   \\
  $3$ & $87.61\%$  & $93.87\%$  & $96.97\%$   \\
  $4$ & -          & $93.63\%$  & $96.90\%$   \\
  $5$ & -          & -          & $96.84\%$
 \end{tabular}
 \caption{\label{tab:p_suc_a} Probability of success of the probabilistic connection operation in a setup where $n$ connection steps are performed prior to purification, with gate error parameter $p$ equal to channel noise $q$.}
\end{table}

\end{document}